\documentclass[twocolumn,english,aps,prl,floatfix,showpacs]{revtex4}
\usepackage[T1]{fontenc}
\usepackage[latin1]{inputenc}
\usepackage{amsmath}
\usepackage{graphicx}
\usepackage{amssymb}

\makeatletter

\usepackage{graphicx}
\usepackage{epsfig}
\usepackage{latexsym}
\usepackage{color}
\usepackage{subfigure}
\usepackage{array}

\usepackage{babel}
\makeatother
\begin{document}

\title{Spin transport and quasi 2D architectures for donor-based quantum computing}

\author{L.C.L. Hollenberg,  A.D. Greentree, A.G. Fowler and C.J. Wellard}

\address{Centre for Quantum Computer Technology
School of Physics, University of Melbourne, VIC 3010, Australia}


\begin{abstract}
Through the introduction of a new electron spin transport
mechanism, a 2D donor electron spin quantum computer architecture
is proposed. This design addresses major technical issues in the
original Kane design, including spatial oscillations in the
exchange coupling strength and cross-talk in gate control. It is
also expected that the introduction of a degree of non-locality in
qubit gates will significantly improve the scaling fault-tolerant
threshold over the nearest-neighbour linear array.
\end{abstract}
\maketitle

The Kane paradigm of donor nuclear spin quantum computing in
silicon \cite{Kane}, based on single atom placement fabrication
techniques \cite{Schofield,Jamieson}, is an important realization
of Feynman's original concept of nanotechnology in the
solid-state. Variations on this theme include electron spin qubits
\cite{Vrijen,DeSousa,Hill} and charge qubits \cite{charge_qubit}.
There are significant advantages of the donor spin as a qubit,
including uniformity of the confinement potential and high number
of gate operations possible within the electron spin coherence
time, measured to be in excess of 60ms \cite{Tyryshkin}.
Consequently, there is great interest in donor-based architectures
and progress towards their fabrication \cite{Clark Review,
Schenkel,Buehler}.

It is often assumed that solid-state designs should be inherently
scalable given the capabilities of semi-conductor device
fabrication. In reality this weak-scalability argument should be
replaced with a stronger version as scalability of a given
architecture is considerably more complex than fabricating many
interacting qubits. Fault-tolerant scale-up requires quantum error
correction over concatenated logical qubits with all the attendant
ancillas, syndrome measurements, and classical feed-forward
processing. Both parallelism and communication must be optimised
\cite{Steane}. Only by considering such systems-level issues in
conjunction with the underlying qubit physics will the
requirements of quantum computation in a given implementation be
understood, and new concepts generated. In this paper we introduce
a new mechanism for coherent donor electron spin state transport,
and in a similar design path to the QCCD ion trap proposal
\cite{Kielpinsky}, we construct a 2D donor architecture based on
distinct qubit storage and interaction regions.

The significant interest in scaling up the donor-based solid-state
designs, has led to a number of works considering these
scalability issues. As a result, several serious problems have
been identified, including: sensitivity of the exchange
interaction and control to qubit placement (at the 2-3 lattice
site level) \cite{KHD,Wellard_Josc,Kane2}, qubit control and
fabrication limitations associated with high gate densities
\cite{Oskin,Copsey1}, spin readout based on spin-charge
transduction \cite{Kane,LHFIR}, and the communication bottlenecks
for linear nearest neighbour (LNN) qubit arrays
\cite{Oskin,Metodiev}.

The issue of local versus non-local fault-tolerant operation is
non-trivial \cite{Gottesman,Svore}. A recent surprising result is
that Shor's algorithm can be implemented on a LNN circuit for the
minimal qubit case with no increase at leading order in the
circuit gate count or depth \cite{FowlerLNNShor,Devitt}. However,
at the systems level one expects a linear nearest neighbour qubit
array to suffer from swap gate overheads, particularly when
concatenated qubit encoding is employed. The general analysis in
 \cite{Svore} shows that locality forces the threshold down
inversely with the physical encoding scale. Recently, the extent
of the LNN penalty has been estimated to bring the threshold down
by two orders of magnitude compared to the non-local case
\cite{Szkopek}.

\begin{figure}[tb]
\centerline{\includegraphics[width=0.9\columnwidth,clip]{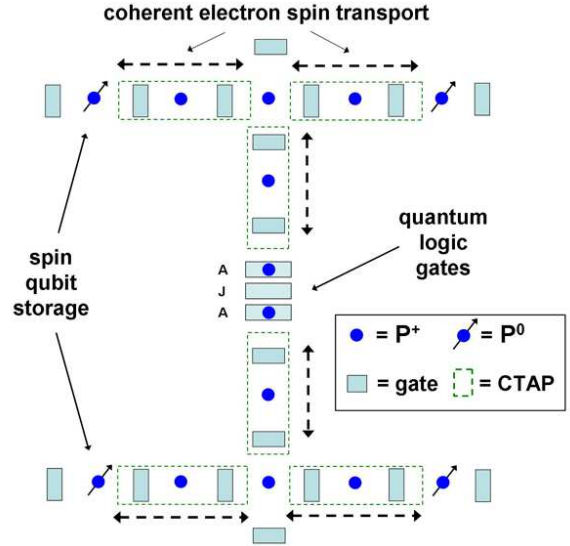}}
\caption{Top view of the 2D donor electron spin quantum
architecture for the case of Si:P, incorporating coherent
transport by adiabatic passage (CTAP). } \label{fig:concept}
\end{figure}

For the Kane, or related donor based architectures, all of the
above implies the imperative of finding ways of traversing the
linear array constraints, as the most effective way to improve the
threshold and tackle the technical problems listed. An important
step in this direction is the proposal for sub-interfacial
transport of electrons in a one dimensional array \cite{Skinner}.
This design has many desirable features, digitising the single and
two qubit gate problems in an elegant way, but also has problems
with scalability due to the relative closeness of gates
\cite{Oskin}.

\begin{figure}[tb]
\includegraphics[width=0.70\columnwidth,clip]{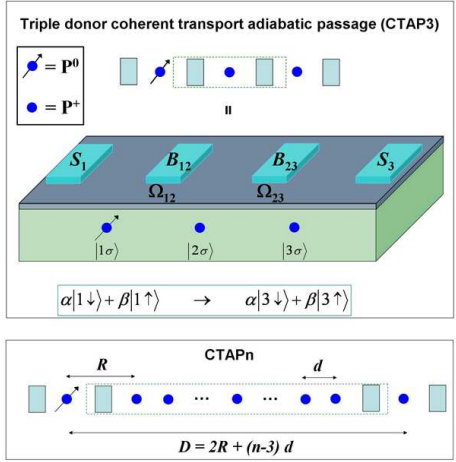}
\caption{\label{fig:TripleWell} Top: Schematic of the one-electron
triple donor system 3D$^{2+}$ based on P donors in silicon. Two of
the donors are assumed ionized, the other neutral. Bottom:
multi-donor CTAPn straddling schemes.}
\end{figure}

The 2D architecture introduced here requires relatively low gate
densities and specifically address the problems listed above. In
Fig. \ref{fig:concept} the geometry is shown for the specific case
of the exchange-interaction based Kane architecture. We note that
the transport ideas presented here allow for a similar, but
non-trivial development for the digital-Kane case. A buried array
of ionised donors provide pathways for coherent transport of
electron spins for in-plane horizontal and vertical shuttling
(dashed-border sections) of qubit states into and out of the
interaction zone. The overall gate density is low compared to the
Kane case, and can be further reduced by increasing the transport
pathway length (Fig. \ref{fig:TripleWell}). Initially all gates
inhibit tunnelling along any given channel. Coherent spin
transport along one segment is achieved by adiabatically lowering
the barriers in a well defined sequence to effect coherent
transfer by adiabatic passage (CTAP) without populating the
intervening channel donors \cite{GreentreeCTAP}. We show that with
appropriate donor separations, the shuttling time can be in the
nanosecond range for one section. In Fig. \ref{fig:concept} the
coherent transport scheme is defined for the minimum number of
donors. Higher order schemes with more donors reduces the gate
density (see Fig. \ref{fig:TripleWell}).

Logic gates are carried out in interaction zones distinct from
qubit storage regions -- shown in Fig. \ref{fig:concept} are the
cannonical A and J gates for electron spin based qubit control at
the microsecond level \cite{Hill}. After mandatory precision
characterisation \cite{SchirmerOi, Cole}, interaction regions with
unacceptably low couplings can be identified and {\em bypassed} in
the circuit flow, thereby avoiding bottleneck issues arising from
the sensitivity of the exchange interaction to donor placement.
This design allows for new variations on the theme, e.g.
digitisation of hyperfine control \cite{Skinner}, or introduction
of local buried B-field antennae structures \cite{LidarBwires},
and space for SET readout techniques
\cite{Kane,LHFIR,GreentreeTriple}.

A schematic of the minimal three donor transport pathway is given
in Fig.~\ref{fig:TripleWell}. The triple-well system
$|1\sigma\rangle$, $|2\sigma\rangle$, $|3\sigma\rangle$ ($\sigma=
\uparrow,\downarrow$) facilitates coherent state transport from
$\alpha|1\downarrow\rangle + \beta|1\uparrow\rangle$ to
$\alpha|3\downarrow\rangle + \beta|3\uparrow\rangle$ {\em without}
populating the $|2\sigma\rangle$ states. Techniques for coherent
transfer by adiabatic passage are well known \cite{Vitanov}, and
for the donor system was proposed in  \cite{GreentreeCTAP} for the
case of charge transfer. A superconducting version of the three
state case has also been proposed \cite{Siewert}. The system is
controlled by shift gates, $S$, which can modify the energy levels
of the end donors, and barrier gates, $B_{i,i+1}$ which control
the tunnelling rate $\Omega_{i,i+1}$ between donors $i$ and $i+1$.

Although the scheme we introduce here necessarily includes spin,
we first consider the zero field case and ignore spin degrees of
freedom \cite{GreentreeCTAP} to illustrate the principles of CTAP
in the one-electron three-donor system, 3D$^{2+}$. The effective
Hamiltonian for the 3D$^{2+}$ system is:
\begin{equation}
\mathcal{H} = \Delta |2\rangle\langle 2| - \hbar \left(
\Omega_{12} |1\rangle\langle 2| + \Omega_{23} |2\rangle\langle 3|
+ {\rm h.c.}\right),
\end{equation}
where $\Omega_{ij}=\Omega_{ij}(t)$ is the coherent tunnelling rate
between donors $|i\rangle$ and $|j\rangle$ and
$\Delta=E_2-E_1=E_2-E_3$. The eigenstates of ${\cal H}$ (with
energies ${\cal E}_{\pm}$ and ${\cal E}_0$) are
\begin{eqnarray}
|{\cal D}_+\rangle &=& \sin \Theta_1 \sin \Theta_2 |1\rangle +
        \cos \Theta_2 |2\rangle +
        \cos \Theta_1 \sin \Theta_2 |3\rangle, \nonumber \\
|{\cal D}_-\rangle &=& \sin \Theta_1 \cos\Theta_2 |1\rangle -
        \sin \Theta_2 |2\rangle +
        \cos \Theta_1 \cos \Theta_2|3\rangle, \nonumber \\
|{\cal D}_0\rangle &=& \cos \Theta_1|1\rangle
        -\sin \Theta_1 |3\rangle,
\label{eq:DressedStates}
\end{eqnarray}
where we have introduced $\Theta_1 = \arctan
\left(\Omega_{12}/\Omega_{23} \right)$ and $\Theta_2 = \arctan
[2\hbar\sqrt{(\Omega_{12})^2 + (\Omega_{23})^2} / \Delta]/2$.
Transfer from state $|1\rangle$ to $|3\rangle$ is achieved by
maintaining the system in state $|{\cal D}_0\rangle$ and changing
the characteristics of $|{\cal D}_0\rangle$ adiabatically ($|{\cal
E}_0 - {\cal E}_{\pm}| \gg |\langle \dot{\cal D}_0|{\cal
D}_{\pm}\rangle|$) from $|1\rangle$ at $t=0$ to $|3\rangle$ at
$t=t_{\max}$ by appropriate control of the tunnelling rates,
without population leakage into the other eigenstates.

\begin{figure}[tb]
\centerline{\includegraphics[width=0.75\columnwidth,clip]
{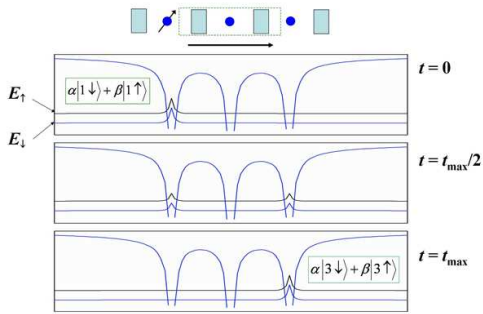}} \caption{Numerical simulation of the CTAP pulse
scheme applied to a spin superposition at donor 1 at $t=0$,
demonstrating coherent transfer to the 3rd donor at $t=t_{\rm
max}$.} \label{fig:CTAP0}
\end{figure}

For the case of coherent spin transport we write the 3D$^{2+}$
Hamiltonian in terms of spin/site operators as:
\begin{equation}
{\cal H} = \sum_{i=1}^{3}
\sum_{\sigma=\uparrow,\downarrow}{E_{i\sigma}}
c^{\dagger}_{i\sigma}c_{i\sigma} + \sum_{<ij>}
\sum_{\sigma=\uparrow,\downarrow}
\Omega_{ij}(t)c^{\dagger}_{j\sigma}c_{i\sigma}
\end{equation}
and numerically solve for the density matrix, $\rho(t)$, in the
presence of a (dominant) charge dephasing rate $\Gamma$, assumed
to act equally on all coherences. Without attempting to fully
optimize control we apply Gaussian pulses of the form $
\Omega_{ij}(t) = \Omega_{ij}^{\max}
    \exp\left[-(t-t_{ij})^2/(2 w^2_{ij})\right]$, where
$t_{ij}$ and $w_{ij}$ are the peak time and width of the control
pulse modulating the tunnelling rate between position states
$|i\rangle$ and $|j\rangle$. To simplify matters for initial
simulations we set the maximum tunnelling rates and standard
deviations for each transition to be equal, i.e.
$\Omega_{ij}^{\max} = \Omega^{\max}$ and $w_{ij}=w$, and set
$\Delta=0$ (these conditions can be relaxed with no effect on the
conclusions of this paper). Transfer is then optimized when the
width of the pulses equals the time delay between the pulses
\cite{Gaubatz}. With total pulse time $t_{\max}$, we choose $w =
t_{max}/8$ so that $t_{12} = (t_{\max} + w)/2$ and $t_{23} =
(t_{max} - w)/2$. This ordering, where $\Omega_{23}$ is applied
\textit{before} $\Omega_{12}$ is known as the counter-intuitive
pulse sequence and has significant advantages in improving
transfer fidelity over other pulse sequences \cite{GreentreeCTAP}.
In Fig.~\ref{fig:CTAP0} we present results showing transport using
the counter-intuitive pulse ordering for a spin superposition
(phases relative to the untransported state).

\begin{figure}[tb]
\centerline{\includegraphics[width=0.95\columnwidth]{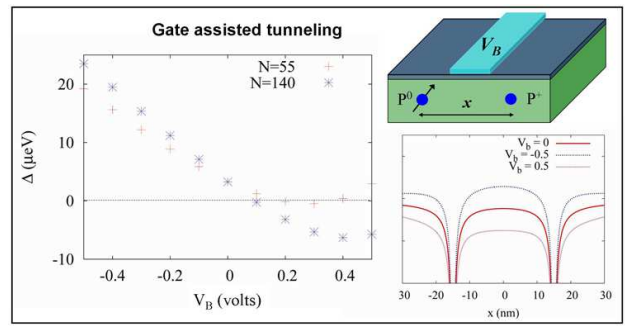}}
\caption{Main: Energy gap for the P-P$^+$ system as a function of
B-gate bias $V_b$ for R=30nm (depth 30nm below interface, 10nm
gate width, basis sizes $N$ = 55 and 140). Lower right: response
of the P-P$^+$ inter-donor potential profile to the barrier gate
bias $V_b = (0,\pm 500)$ mV.}\label{fig:potential}
\end{figure}

Generally, when the adiabaticity criterion is satisfied and the
transport time is at least an order of magnitude faster than
charge dephasing, the transport fidelity is high. These results
are consistent with those of Ivanov et al \cite{Ivanov} who
considered the role of dephasing in three-state Stimulated Raman
Adiabatic Passage (STIRAP). Although these competing timescales
are essentially unmeasured at present, estimates
\cite{Barrett,Fedichkin} for the P-P$^+$ charge dephasing time are
of order 10ns and a value of 220ns was reported recently for a
Si:P double-dot \cite{Gorman}, whereas sub-nanosecond tunnelling
times are possible due to the strong confining potential of donor
nuclei. The CTAP transport time will be defined primarily by the
gate-assisted tunnelling rate, which we calculate as follows.
Using the TCAD package we compute the potential due to a surface
B-gate bias and determine the donor electron wave function in an
effective mass basis, e.g. $F^{n,l,m}_{\pm z}({\bf r}) =
\varphi_{n,l,m}(x,y,\gamma z)$, about the six band minima where
the $\varphi_{n,l,m}$ are hydrogenic orbitals with Bohr radius
$a_\perp$, and $\gamma = a_\perp/a_\parallel$. Diagonalising the
total Hamiltonian of the system, using pseudopotentials to
describe the silicon bandstructure, we obtain a generalised
Kohn-Luttinger wave function:
\begin{equation}
\psi({\bf r},V) = \sum_{n,l,m} c_{n,l,m}(V) \sum_{\mu=1}^6
F_\mu^{n,l,m}({\bf r}) {\rm e}^{i {\bf k}_\mu .{\bf r}} u_{{\bf
k}_\mu} ({\bf r}), \label{equation:donor_wf(v)}
\end{equation}
where the Bloch states are $u_{{\bf k}_\mu} ({\bf r})= \sum_{\bf
G} A_{{\bf k}_\mu}({\bf G}) e^{i {\bf G}\cdot {\bf k}_\mu}$. We
form bonding and anti-bonding states $\Psi_\pm({\bf r},V) = {\cal
N}(\psi_L({\bf r},V) \pm \psi_R({\bf r},V))$, normalised by ${\cal
N}$, and compute the gap as shown in Fig. \ref{fig:potential} for
basis sizes 55 and 140 ($n_{\rm max}$ = 5 and 7). Comparison of
the non-linear regions indicates that the range of validity is
$|V_b|\lesssim $ 200 mV.

In contrast to what one expects for an isolated P-P$^+$ system in
vacuum where the nodal structure of the bonding and anti-bonding
states is simple, the non-trivial nodal properties of the donor
electron wave function and the proximity of the oxide interface
complicates the tunnelling control. These calculations directly
extend similar effects noted in the ungated P-P$^+$ system
\cite{Hu_Charge_Qubit}. From Fig. \ref{fig:potential} we see that
for this configuration the tunnelling rate can be varied from zero
at +100mV to $\sim$ 10 GHz at -200mV, giving a gate assisted
tunnelling time of 60 ps.

\begin{figure}[tb]
\centerline{\includegraphics[width=0.8\columnwidth,clip]
{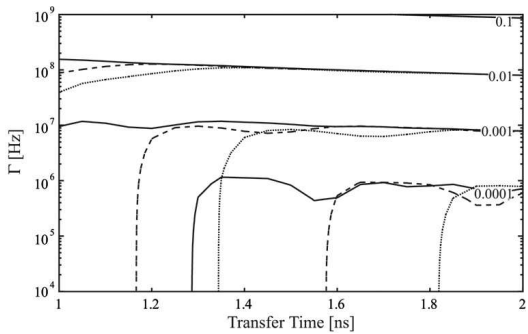}} \caption{Transfer error as a function of charge
dephasing rate and total transfer time for CTAP5 (solid line),
CTAP7 (long dashes) and CTAP9 (short dashes) for the case of 30nm
end-donor spacings, and 20 nm between the central
donors.}\label{fig:CTAP1}
\end{figure}

Based on this value, CTAP simulations for 5, 7 and 9 donor chains
are presented in Fig.~\ref{fig:CTAP1}. The adiabatic nature of the
transport scheme provides an inherent robustness, as evidenced in
Fig.~\ref{fig:CTAP1}, which shows a remarkable uniformity in the
response to charge dephasing for the different path lengths once
the adiabatic regime is reached. Another consequence is that
inevitable variations in tunnelling rates due to donor placement
\cite{Hu_Charge_Qubit} will not affect the viability of the
scheme, as further simulations have explicitly verified. The
extent to which $\Gamma$ controls the transport fidelity is also
clear, although we note that there is room for improvement through
optimisation of control pulses and minimisation of charge
fluctuations through fabrication development. Non-zero transport
errors may require monitoring mechanisms for heralding successful
transport, or an error correction protocol for transport loss. As
intrinsic spin-orbit coupling for donor states in silicon is very
low, dephasing of donor electron spin is dominated by spectral
diffusion due to spin impurities and is mitigated by isotopic
purification \cite{DeSousa2}. For the bound state spin-orbit
coupling, at $V_B\sim 200$ mV we calculate from
Eqn(\ref{equation:donor_wf(v)}) the non-S components to be
$\sum_{n,l>0,m} |c_{n,l,m}(V)|^2 < 10^{-4}$ indicating that the
deviation from the S sector is minimal. Together with the zero
occupation of channel states, this suggests that charge dephasing
will have a negligible second order effect on the spin coherence
during transport. Decoupling of orbital and spin sectors has
already given rise to demonstrations of coherent transport of
electron spins over 100$\mu$m \cite{Kikkawa}.

The basic layout of 2D donor arrays, with storage regions,
vertical and horizontal transport pathways and interaction zones,
allows us to explicitly consider designs for fault-tolerant
operation. For example, we can arrange the logical qubit groups
and ancillas so that the transport rails allow for non-local
intralogical qubit interactions (qubit -- ancillas) and LNN
interlogical interactions. With inherent parallelism of operation,
interlogical gates can then be applied transversally as required
to implement fault-tolerant gates. Another possibility with less
stringent fabrication requirements is a linear qubit storage with
transport and interaction rails either side.

The optimum arrangement for fault-tolerant operation requires
sophisticated systems level simulations \cite{Copsey2} to
determine the best use of this medium range quantum transport
capability, and the corresponding improvements on the LNN
threshold. In any case, it is clear that the introduction of
coherent spin transport to donor quantum computing allows us to
address many problems in the Kane concept, and consider scalable
fault-tolerant architectures with low gate densities, room for SET
structures and control, and a bypass mechanism for low value
exchange gates. One expects the realities of the silicon
crystaline environment will necessitate the characterisation of
transport pathways, however, the precision requirements of the
adiabatic CTAP mechanism would be far less than the quantum gate
threshold.

We thank S. Das Sarma, G. Milburn, F. Wilhelm and J. Cole for
comments, and support of the Australian Research Council, the U.S.
National Security Agency, Advanced Research and Development
Activity and Army Research Office under contract DAAD19-01-1-0653.

\end{document}